\documentclass[onecolumn]{aastex7}

\usepackage{graphicx}   
\usepackage{subcaption} 
\usepackage{amsmath}  
\usepackage{float}
\usepackage{wrapfig}
\usepackage{xcolor}

\newcommand{\eg}{e.g.,\ }

\begin{document}

\title{Formation Pathways of Compact Elliptical Galaxies: Perspective of Internal Structures}

\correspondingauthor{Soo-Chang Rey}  
\email{screy@cnu.ac.kr}              

\author{Jinwon Yi}
\affiliation{Department of Astronomy and Space Science, Chungnam National University, Daejeon 34134, Republic of Korea}
\email[]{jinwon1243@gmail.com}

\author[orcid=0000-0002-0041-6490]{Soo-Chang Rey} 
\affiliation{Department of Astronomy and Space Science, Chungnam National University, Daejeon 34134, Republic of Korea}
\email[show]{screy@cnu.ac.kr}

\author[orcid=0000-0003-3474-9047]{Suk Kim}
\affiliation{Department of Astronomy and Space Science \& Research Institute of Natural Sciences, Chungnam National University, Daejeon 34134, Republic of Korea}
\email[]{star4citizen@gmail.com}

\author[orcid=0000-0002-6261-1531]{Youngdae Lee}
\affiliation{Department of Astronomy and Space Science \& Research Institute of Natural Sciences, Chungnam National University, Daejeon 34134, Republic of Korea}
\email[]{hippo206@gmail.com}

\begin{abstract}

We present a structural analysis of 138 compact elliptical galaxies (cEs) in the redshift range of $z < 0.05$ using the Sloan Digital Sky Survey (SDSS) DR12 data. We perform single- and double-component Sérsic model fitting to their SDSS $r$-band surface brightness profiles. By dividing cEs into those with [cE(w)] and without [cE(w/o)] a bright host galaxy, we find a significant structural dichotomy: the majority ($\sim$85\%) of cE(w)s exhibit single-component profiles, while a similar proportion ($\sim$85\%) of isolated cE(w/o)s display double-component profiles, characterized by a compact, inner component and a diffuse, disk-like outer component. These results suggest that host-associated cE(w)s primarily form through the tidal stripping of larger progenitors, resulting in a compact bulge-like core. In contrast, isolated cE(w/o)s appear to form intrinsically at early epochs, likely through gas-rich mergers, and retain disk-like outer structures. The Sérsic index distribution of cE(w)s with single-component structure indicates progenitor types ranging from pseudo-bulge to classical bulge, supported by differences in stellar populations. A small fraction of cEs, including double-component cE(w)s and single-component cE(w/o)s, suggests complex evolutionary channels involving environmental capture or ejection. Our results emphasize that the structural characteristics of cEs, specifically the presence or absence of an extended outer envelope, serve as a crucial diagnostic tool to distinguish tidally stripped remnants from intrinsically formed low-mass cEs in isolation.

\end{abstract}

\bigskip 

\section{Introduction}
\label{sec:intro}

Compact elliptical galaxies (cEs) are rare galaxies in the local universe, characterized by their small size and high stellar density. With stellar masses, spanning from $10^8$ to $10^{10}\,\mathrm{M_\odot}$ and their effective radii smaller than 1 kpc, these systems occupy a distinct distribution in the size vs. stellar mass plane, differentiating them from early-type dwarf galaxies while following the relation of massive early-type galaxies \citep{Norris14, Ferre-Mateu18, Kim20}. The majority of cEs also deviate significantly from the conventional mass--metallicity relation observed in early-type galaxies, exhibiting higher metallicities than galaxies of comparable mass \citep{Janz16, Kim20, Ferre-Mateu21a}. Their high stellar metallicity and compact size render them anomalous within a broad galaxy population.

cEs are predominantly found in dense environments such as galaxy clusters and groups \citep{Norris14, Chilingarian15, Janz16, Kim20}. Furthermore, they are often found to be in proximity to massive host galaxies, suggesting a physical connection between them \citep{Zhang17, Kim20, Chen22}. However, an increasing number of cEs have also been discovered in isolated environments with no massive galaxies nearby \citep{Huxor13, Kim20, Chen22}. In fact, some cosmological simulations produce most (\,$\sim$70\%) of their cEs in isolation (\eg\ \citealt{Deeley23}). The ubiquity of cEs across a diverse range of global (cluster, group, and field) and local (host-associated and isolated) environments suggests that multiple formation pathways may be responsible for their existence.

Two primary formation mechanisms have been proposed to account for the unique properties of the cEs. The association with an adjacent massive host galaxy suggests a tidal stripping scenario in which cEs are remnants of larger, more massive galaxies (\eg\ compact bulges of early-type disk galaxies) that have undergone significant stripping of their outer stellar envelopes owing to interactions with massive hosts \citep{Faber73, Bekki01, Choi02, Graham02, Chilingarian09a}. The result of this process is the formation of metal-rich, low-mass cEs, which corroborates the observational results of the mass--metallicity relation \citep{Norris14, Janz16, Chilingarian15, Kim20, Ferre-Mateu18, Ferre-Mateu21a}. Further evidence supporting cE formation through the stripping process includes the detection of overmassive black holes (BHs) in some cEs, which show larger BH masses than expected for galaxies with comparable stellar mass or velocity dispersion (\citealt{Forbes14, Pechetti17, Ferre-Mateu21b}). Moreover, the discovery of tidal streams around a few cEs provides direct evidence of this mechanism \citep{Huxor11, Paudel13, Chilingarian15, Ferre-Mateu18}. 

The presence of cEs in isolated environments \citep{Huxor13, Paudel14, Kim20, Chen22} suggests an alternative formation scenario, in which they are natural, intrinsically compact objects \citep{Wirth84, Kormendy09, Kormendy12}. This proposition is supported by the result that isolated cEs follow the mass--metallicity relation of massive early-type galaxies in their low-mass range \citep{Kim20, Ferre-Mateu21a}. The observational properties of the central BHs of the isolated cEs also confirm that they are bona fide low-mass compact galaxies \citep{Paudel16, Rey21, Ferre-Mateu21b}. It has been suggested that isolated cEs might have formed at high redshifts and survived into the local universe by lack of mergers with other galaxies \citep{Zolotov15, Wellons16, Deeley23}. 

Studies on the structural properties of galaxies contribute to our understanding of the formation and the underlying physical processes that govern galaxy evolution in different environments. Dense environments accelerate galaxy transformations through various environmental processes (\eg\ \citealt{Dressler80}). Comparative studies of isolated galaxies have provided insights into intrinsic evolution without external perturbations \citep{Geha12}. cEs are of particular interest in this context, because they exhibit high surface brightness and stellar density, similar to the cores of more massive galaxies rather than typical low-mass galaxies \citep{Faber97, Zhang17, Ferre-Mateu18, Ferre-Mateu21a}. To robustly test the formation scenarios of cEs and understand the role of different environments, studying the structural properties of large cE samples alongside various physical properties is essential.

If cEs are formed from the tidal stripping of larger, more massive progenitors through interactions with massive neighboring host galaxies, this process would strip away the outer stellar component of progenitors and leave behind a dense, compact core \citep{Bekki01, Chilingarian09a}. In this scenario, the cEs are expected to possess a simple structure with a single core component. In contrast, if cEs are intrinsic low-mass compact galaxies in isolation formed from a hierarchical merging process in the early universe without any tidal stripping, they would be expected to feature more complex structures similar to massive compact passive galaxies at high redshifts (\eg\ \citealt{Toft05, Buitrago08, van11, Chang13, Trujillo06, Trujillo14, Zolotov15}). Depending on the different environments, internal structures of the cEs may provide crucial information that distinguishes between competing formation pathways. 

However, the structures of cEs have not been comprehensively investigated for extensive samples in previous studies. Prior studies on the structures of cEs have focused on a limited number of these galaxies (\eg\ \citealt{Chilingarian07, SmithCastelli08, Price09, Chilingarian10, Paudel14, Guerou15, Ferre-Mateu21a, Caso24}). Furthermore, numerous studies have concentrated on specific environments, such as clusters, without considering a diverse range of environments \citep{Chilingarian07, SmithCastelli08, Price09, Caso24}. 

\citet{Kim20} examined the stellar population properties of 138 cEs according to their small-scale, local environment (\eg with a massive host galaxy or isolated) and global environment (cluster, group, and field), providing insights into how the environment influences cE formation (see also \citealt{Chen22}). They identified two distinct formation channels: tidal stripping for cEs associated with a massive host and in-situ evolution for isolated cEs. In this study, we aim to investigate the internal structures of local cEs in relation to their local environment by utilizing the large and diverse sample of \citet{Kim20}. This approach allows us to examine whether the structures of cEs can elucidate their formation mechanisms and evolutionary pathways between host-associated and isolated cEs. In Section~\ref{sec:data}, we present a sample of cEs and an analysis of their structures. Section~\ref{sec:result, discussion} presents the structural parameters of cEs depending on their local environment. The formation pathways of the cEs based on their structural properties are also discussed. We conclude with a summary in Section~\ref{sec:summary, conclusion}.

\section{DATA and ANALYSIS}
\label{sec:data}

In our analysis, we utilized a catalog of 138 cEs, characterized by low luminosities ($M_g > -18.7$ mag) and compact sizes, with $r$-band effective radius ($R_{\mathrm{eff}}$) less than 600 pc in the local volume ($z < 0.05$) constructed using the Sloan Digital Sky Survey (SDSS) DR12 (see \citealt{Kim20} for details on the selection of cEs). \citet{Kim20} derived the age, [Z/H], and [$\alpha$/Fe] values of the cEs by comparing the Lick indices obtained from SDSS spectra with simple stellar population model grids. Additionally, they calculated the stellar masses of the cEs using the relation between the SDSS $g-r$ color and the stellar mass-to-light ratio based on the $i$-band luminosity \citep{Bell03}. The stellar masses of the cEs span the range of $\log(M_\ast / M_\odot) = 8.3$ -- $9.5$.

\citet{Kim20} assessed the small-scale, local environment of the cE sample by measuring the projected distances from each cE to the nearest luminous ($M_r < -21.0$ mag) host galaxies with velocity differences less than 500 km s$^{-1}$. The virial radius ($R_{\mathrm{vir}}$) of each host galaxy was estimated using the relation $R_{\mathrm{eff}} = 0.015\,R_{\mathrm{vir}}$ \citep{Kravtsov13}, where $R_{\mathrm{eff}}$ of the host galaxy was taken from the SDSS DR12. The cEs are subsequently categorized into two groups, cE(w)s and cE(w/o)s, based on their proximity to the host galaxy, with cE(w)s located within one $R_{\mathrm{vir}}$ of the host galaxy and cE(w/o)s situated outside this radius. While cE(w)s are physically associated with the host galaxy, the majority of cE(w/o)s are predominantly situated in low-density, isolated local environment akin to those of star-forming dwarf galaxies (see \citealt{Kim20} for details). The final samples comprise 65 cE(w)s and 73 cE(w/o)s. \citet{Kim20} also estimated the global environment, encompassing cluster, group, and field, of cEs utilizing the group catalog of \citet{Tempel14}. Within the cE sample, 23, 24, and 91 cEs are identified in the cluster, group, and field, respectively. We adopted the stellar population properties and environment of the cEs from \citet{Kim20}.

To perform surface photometry of the cEs, we utilized sky-subtracted $r$-band images from the SDSS DR12. We manually masked all background and foreground objects surrounding the cEs. For bright galaxies, such as host galaxies near the cE, GALFIT \citep{Peng02} was used to subtract their light. The surface brightness profiles (SBPs) of all the cEs were measured using the \texttt{Ellipse} package\footnote{\texttt{photutils.isophote} is part of the \textit{Photutils} package \citep{photutils}.}. During ellipse fitting, the center, ellipticity, and position angle were allowed to vary. We performed single- and double-component Sérsic model fitting for all cEs using the Levenberg-Marquardt least-square method\footnote{\texttt{astropy.modeling.fitting.LevMarLSQFitter} is part of the \textit{Astropy} package \citep{astropy:2022}.}. The central region within 1.32 arcseconds of the SBPs was excluded from the model fitting to avoid the seeing effect. The value of 1.32 arcseconds corresponds to the median FWHM of the SDSS DR12 $r$-band imaging \citep{York00}. Furthermore, only SBPs brighter than $27$ mag arcsec$^{-2}$ were considered to minimize background contamination \citep{Pohlen06, Trujillo06}.

We determined the best-fit model for each cE by evaluating the reduced chi-square values for both single ($\chi^2_1$) and double Sérsic functions ($\chi^2_2$). We classified cEs as having single- or double-component SBPs based on the criterion $\Delta\chi^2 \leq 0$ and $\Delta\chi^2 > 0$, respectively, where $\Delta\chi^2$ is defined as $\chi^2_1 - \chi^2_2$. Furthermore, we verified the final classification through a visual examination of the SBPs and the overall residuals, representing the discrepancies between the best-fitting model and the actual data points. Subsequently, we obtained the structural parameters of the cEs, including the Sérsic index ($n$), $R_{\mathrm{eff}}$, and surface brightness at the effective radius ($\mu_{\mathrm{eff}}$).

To ensure the robustness of our analysis, we employed archival $r$-band images from the Hyper Suprime-Cam (HSC) of the 8.2-m Subaru telescope, which provide better data quality in the faint, outer regions of cEs compared to the SDSS images \citep{Aihara22}. By applying the same procedures used for the SDSS images, we obtained the SBPs of 12 cEs and conducted model fitting. The SBPs derived from the HSC data exhibit reduced scatter in the outer regions. Nevertheless, we observed that the SBPs obtained from both the HSC and SDSS data conform to the same model profiles for all cEs. Additionally, the structural parameters derived from the HSC data show excellent agreement with those obtained from the SDSS data: $\Delta n = 0.08 \pm 1.10$, $\Delta R_{\mathrm{eff}}$ (kpc) $= -0.01 \pm 0.29$, and $\Delta \mu_{\mathrm{eff}}$ (mag arcsec$^{-2}$) $= 0.00 \pm 0.46$, where $\Delta$ represents the difference between SDSS and HSC values. These results support the conclusion that our analysis using the SDSS data is sufficiently reliable.

\begin{figure}[h]
\includegraphics[width = \textwidth]{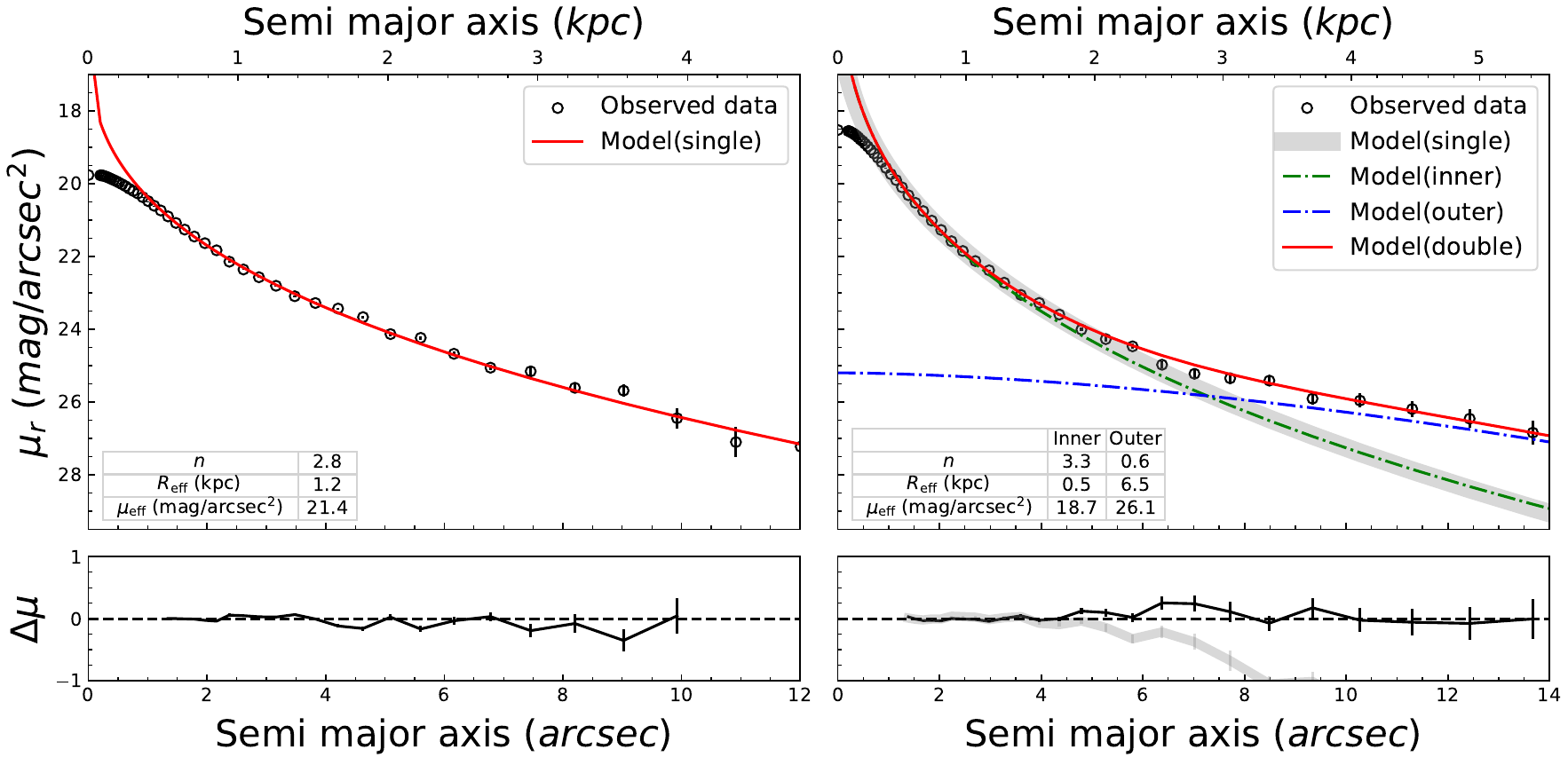}
\caption{Examples of SDSS r-band SBPs (open circles, upper panels) as a function of semi-major axis for single- (left panel) and double-component (right panel) cEs. The red curves denote the optimal fit for the single- (left panel) and double-components (right panel) of the double Sérsic model. The green and blue dot-dashed curves in the right panel represent the inner and outer components of the double Sérsic model, respectively. The gray curve corresponds to the single Sérsic model. The structural parameters derived from the best fit, specifically the Sérsic index ($n$), effective radius ($R_{\mathrm{eff}}$), and surface brightness at the effective radius ($\mu_{\mathrm{eff}}$), are presented in the boxes at the bottom-left of each panel. The lower panels display the residuals, representing the differences between the best-fitting model and the data points.}
\label{fig:1}
\end{figure}

Figure~\ref{fig:1} presents examples of SBPs (upper panels) and residuals (bottom panels) as a function of the semi-major axis (SMA) for single- (left panels) and double-component (right panels) cEs. The single-component cE aligns well with the single Sérsic model (red curve) across the entire SMA range. In contrast, for the double-component cE, the double Sérsic model (red curve) provides a superior fit compared to the single Sérsic model (gray curve). The inner (green dot-dashed curve) and outer (blue dot-dashed curve) components of the double Sérsic model effectively represent the central and outer regions (SMA $> 8$ arcsec, $> 3$ kpc) of the cE, respectively. The double-component cE possesses a faint outer stellar component with a nearly constant $r$-band surface brightness of $\sim$26 mag arcsec$^{-2}$ extending to 6 kpc from its center.

\section{RESULT and DISCUSSION}
\label{sec:result, discussion}

\subsection{Structural Dichotomy of Compact Elliptical Galaxies}

\begin{figure}[h]
\centering
\centering
\includegraphics[width=0.45\textwidth]{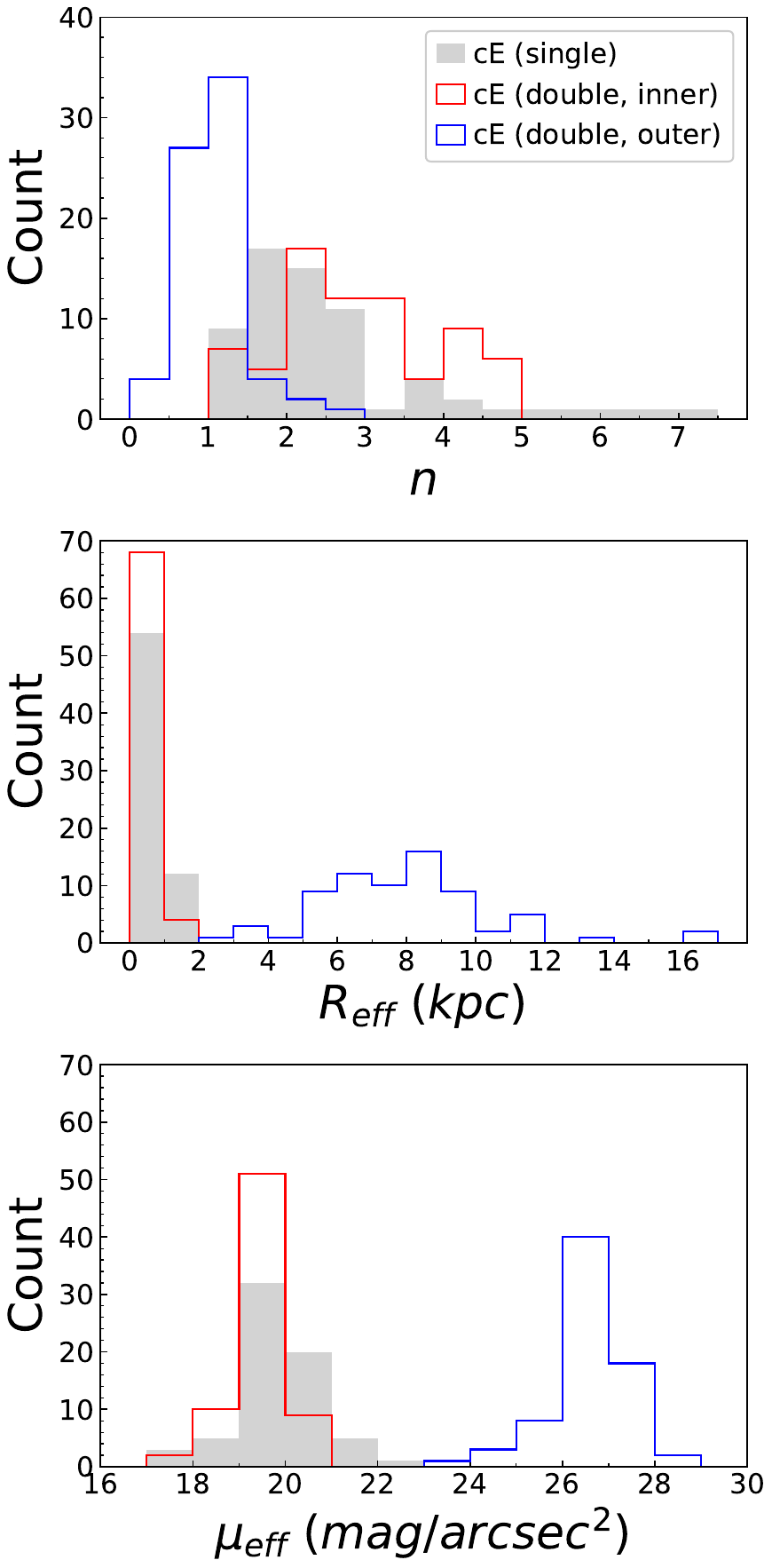}
\caption{Distributions of structural parameters for single- (filled gray histograms) and double-component cEs (red and blue histograms for inner and outer components, respectively). The top, middle, and bottom panels display the distributions of $n$, $R_{\mathrm{eff}}$, and $\mu_{\mathrm{eff}}$, respectively. Notably, all structural parameters exhibit clear distinctions between the inner and outer components.}
\label{fig:2}
\end{figure}

In Figure~\ref{fig:2}, we present the structural parameters of cEs, dividing them into 66 single-component cEs (filled gray histograms) and 72 double-component cEs (red and blue histograms for the inner and outer components, respectively). A notable characteristic of double-component cEs is the distinctly different distributions of structural parameters between the inner and outer components. The Sérsic index distributions (top panel) of the inner and outer components are clearly separated at \( n \sim 2 \), with the outer components exhibiting lower Sérsic index values (median value of 1.1) compared to the inner components (median value of 2.8). The Sérsic index quantifies the light concentration in galaxies, providing information about their morphological characteristics. This suggests that the outer component possesses a disk-like structure, whereas the inner component displays a centrally concentrated light distribution \citep{Graham05}.

In the middle and bottom panels, the inner and outer components demonstrate distinct distributions of the \( R_\mathrm{eff} \) and \( \mu_\mathrm{eff} \). The \( R_\mathrm{eff} \) of the outer component exhibits a broad distribution, extending to larger radii, while a pronounced peak at small radii characterizes that of the inner component. The \( R_\mathrm{eff} \) of the outer component (median value of 7.9~kpc) is more than ten times larger than that of the inner component (median value of 0.6~kpc). Conversely, the outer component exhibits a fainter surface brightness than the inner component. The outer component (median value of 26.6~mag~arcsec\(^{-2}\)) is over 900 times fainter than the inner component (median value of 19.2~mag~arcsec\(^{-2}\)). These distributions indicate structural differences between the inner and outer components of the double-component cEs, particularly in terms of light concentration, size, and surface brightness.

In the case of single-component cEs, the Sérsic index distribution appears to shift slightly toward lower values, with a median value of 2.3. However, the \( R_\mathrm{eff} \) and \( \mu_\mathrm{eff} \) distributions closely resemble those of the inner component of double-component cEs, exhibiting similar peaks (median values of \( R_\mathrm{eff} = 0.7 \)~kpc and \( \mu_\mathrm{eff} = 19.7 \)~mag~arcsec\(^{-2}\)). Although double-component cEs have a substantial outer component, due to their very faint surface brightness, the apparent size of the cEs remains by selection within a similar range regardless of their detailed structure (e.g., see Figure~4 of \citealt{Kim20}).

\begin{figure}[h]
\centering
\centering
\includegraphics[width=\textwidth]{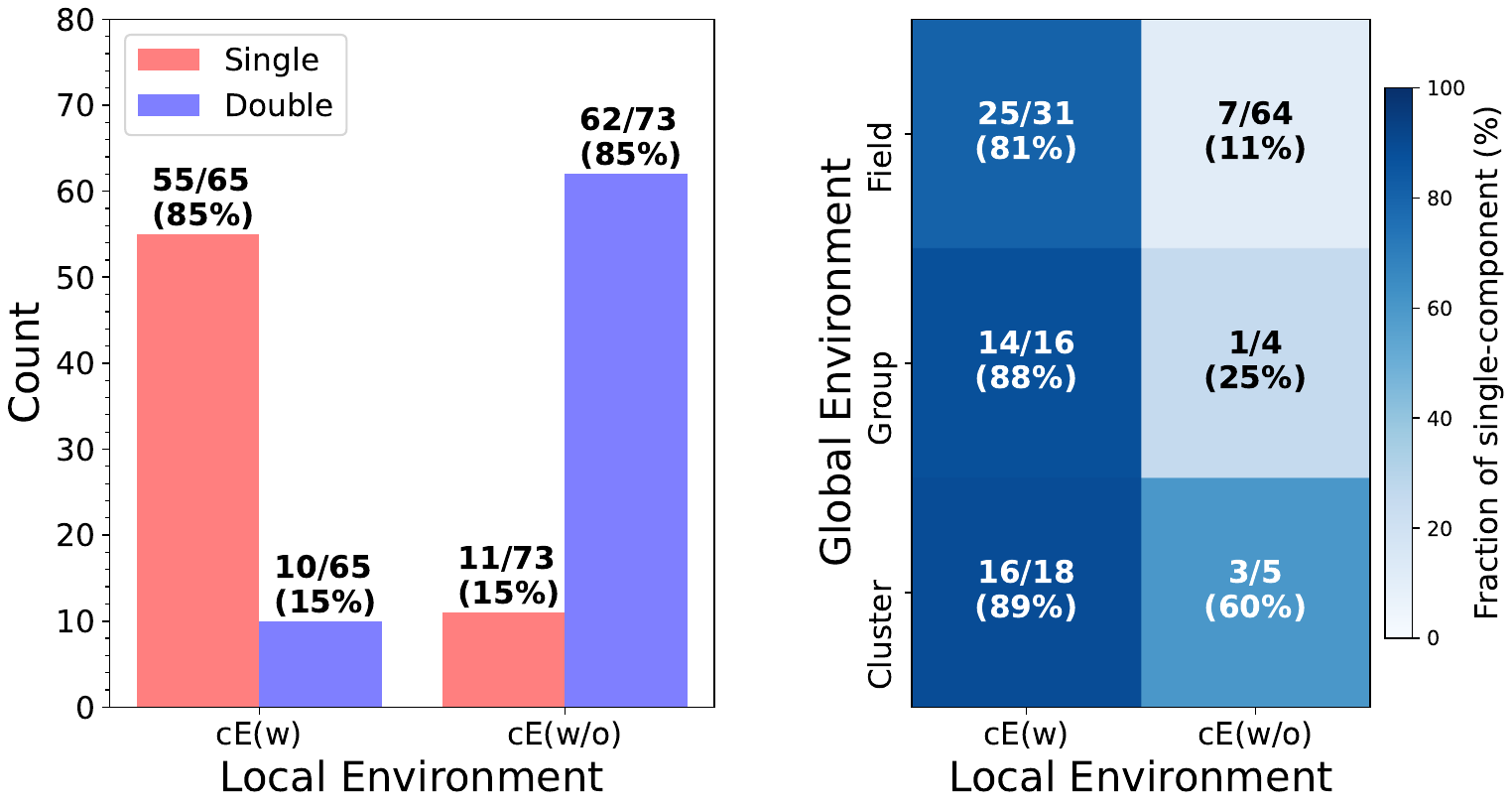}
\caption{(Left) Structural classifications of cEs as a function of their local environment. cE(w) and cE(w/o) denote cEs with and without a nearby massive host galaxy, respectively. Red and blue bars represent single- and double-component cEs, respectively. The numbers above the bars indicate the counts and fractions of cEs in each structural category. Most cE(w)s (85\%, 55/65) display single-component structures, whereas the majority of cE(w/o)s (85\%, 62/73) exhibit double-component structures, highlighting a pronounced dichotomy influenced by local environment. (Right) Matrix illustrating the fraction of single-component cEs across both local (cE(w) and cE(w/o)) and global (cluster, group, and field) environments. Each cell is color-coded and labeled with the percentage and number of single-component cEs relative to the total number of cEs in that category.}
\label{fig:3}
\end{figure}

Figure~3 illustrates the structural classifications of cEs in relation to their local and global environments. The left panel displays the relative fractions of single- and double-component structures as a function of the local environment, specifically differentiating between cE(w) and cE(w/o)s. A notable structural dichotomy is evident: most cE(w)s ($\sim85\%$, 55/65) exhibit single-component structures, whereas the majority of cE(w/o)s ($\sim85\%$, 62/73) show double-component structures. The predominance of simple structures in cE(w)s suggests a strong environmental influence on their structural evolution. Gravitational interactions and tidal forces exerted by host galaxies may transform complex structures into simpler ones and/or inhibit the development of complex structures. By contrast, the prevalence of double-component structures in cE(w/o)s suggests that cEs in isolated environments tend to develop more intricate structures. 

We examined whether the global environment (i.e., cluster, group, and field) exerts an additional influence on cE structure beyond the effect of a nearby massive host galaxy. The right panel of Figure~3 illustrates a matrix that compares the fraction of single-component cEs across both local and global environments. Each cell is color-coded and labeled with the fraction and number of single-component cEs relative to the total number of cEs in that category. We find that the structural properties of cE(w)s remain consistent across various global environments, with single-component fractions of 81\% in the field, 88\% in group, and 89\% in cluster. This consistency suggests that the dominant factor influencing cE(w) structure is the presence of a nearby host galaxy, rather than the large-scale environment. For cE(w/o)s, there is a modest increase in the single-component fraction from the field (11\%) to group (25\%) and cluster (60\%). However, due to the very small sample sizes of cE(w/o)s in group ($n = 4$) and cluster ($n = 5$), these variations lack statistical significance and should be interpreted with caution. Consequently, our results suggest that the local environment, particularly the presence or absence of a massive host galaxy, is the primary factor of structural differences in cEs.

\subsection{Progenitors of Stripped cEs from Sérsic Indices}

The Sérsic profile decomposition applied to most cE(w/o)s with double components posits that the inner component represents the bulge, whereas the outer component corresponds to the disk \citep{Graham02, Paudel16}. In contrast, for most cE(w)s, where a single-component Sérsic model fitting is more suitable, the derived structural parameters can reflect the overall bulge properties. By comparing the structural parameters of a single component of cE(w)s with the inner component of cE(w/o)s, it is possible to investigate any significant differences between cE(w) and cE(w/o) populations, which could provide insights into their formation and evolutionary processes.

\begin{figure}[h]
\centering
\centering
\includegraphics[width=0.45\textwidth]{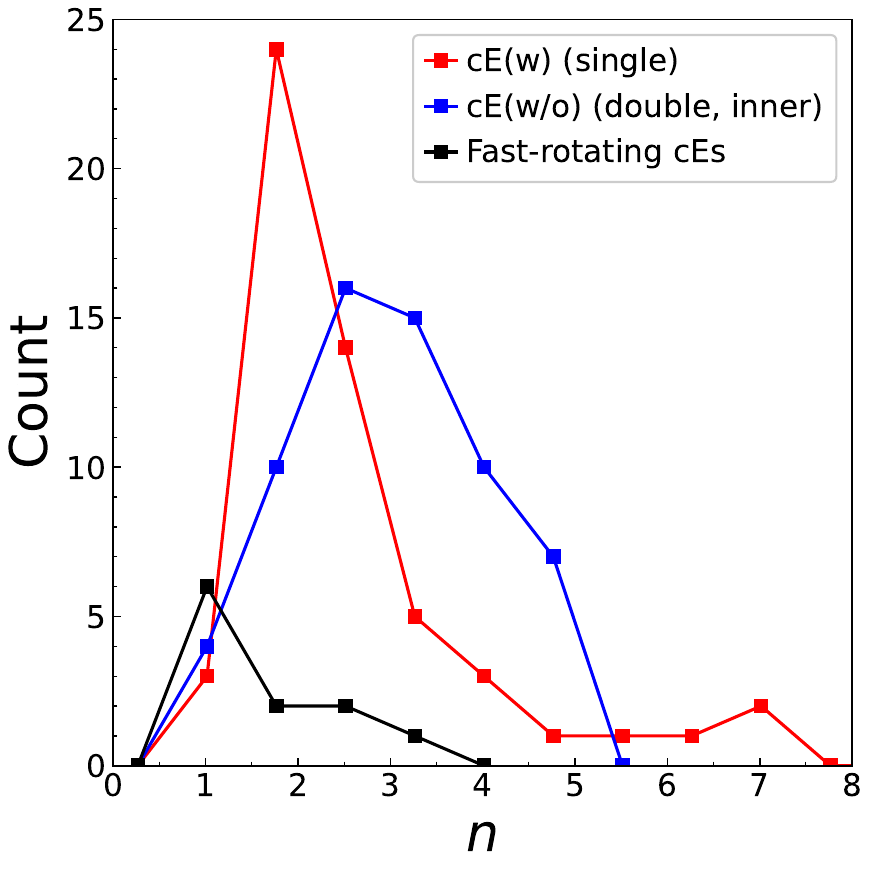}
\caption{Sérsic index distribution of 55 cE(w)s (red curve) and the inner component of 62 cE(w/o)s (blue curve), where the cE(w)s and cE(w/o)s are well fitted with single- and double-component Sérsic models, respectively. The black curve denotes the 11 fast-rotating cEs identified in previous studies \citep{Guerou15, Ferre-Mateu21b}.}
\label{fig:4}
\end{figure}

In Figure~\ref{fig:4}, we present the distribution of the Sérsic index for 55 cE(w)s (red curve) and the inner component of 62 cE(w/o)s (blue curve), where the cE(w)s and cE(w/o)s are well-fitted with single- and double-component Sérsic models, respectively. The distribution of cE(w)s appears to shift toward a lower Sérsic index with a median value of 2.2. In contrast, the inner component of cE(w/o)s exhibit a slightly higher peak and larger dispersion toward higher values, with a median value of 3.0. The higher Sérsic index values of cE(w/o)s are consistent with those of early-type galaxies \citep[e.g.,][]{Kormendy96, Kormendy09}. The distinction between the cE(w)s and inner component of cE(w/o)s suggests differences in their structural characteristics, indicating that cE(w)s have a relatively lower central concentration than the inner component of cE(w/o)s. A two-sample Kolmogorov--Smirnov test confirms that the difference between the two distributions is statistically significant with a $p$-value of 0.001.

Sérsic index is a crucial parameter for differentiating between classical and pseudo-bulges in spiral galaxies \citep{Kormendy04, Fisher08}. Typically, a threshold Sérsic index value of $n \sim 2$ is employed, where pseudo-bulges are characterized by $n < 2$ and classical bulges by $n > 2$ \citep{Fisher08}. Interestingly, a significant fraction (42\%) of the cE(w)s have Sérsic index values below 2. Pseudo-bulges with lower Sérsic indices are often linked to enhanced rotational support \citep{Fabricius12, Hu24b}. Previous studies have identified 11 fast-rotating cEs, primarily associated with host galaxies, that exhibit low Sérsic index values \citep[e.g.,][]{Guerou15, Ferre-Mateu21b}. Figure~\ref{fig:4} illustrates that the majority of these fast-rotating cEs (see black curve, median Sérsic index of 1.2) coincide well with the low Sérsic index values observed in the cE(w) distribution. \citet{Ferre-Mateu21a} proposed, based on the comparison of internal kinematics between cEs and various massive galaxies, that the progenitors of these cEs are likely spiral galaxies. Furthermore, the central pseudo-bulge of the lenticular galaxy NGC 6340 displays characteristics akin to those of cEs within the $k$-space fundamental plane \citep{Chilingarian09b}.

In addition to the structural distinction, it has been proposed that the stellar population properties of pseudo-bulges differ from those of classical bulges and ellipticals \citep[and references therein]{Hu24a, Hu24b}. Classical bulges predominantly compose old, metal-poor stars, whereas pseudo-bulges generally consist of younger, metal-rich stellar populations. This difference is often attributed to the divergent formation mechanisms and star formation (SF) histories: classical bulges are linked to intense SF episodes in the early universe, while pseudo-bulges are associated with secular evolution that supports sustained SF. In this context, we examined the stellar population properties (age, [Z/H], and [$\alpha$/Fe]) of cE(w)s with low ($n < 2$) and high ($n > 2$) Sérsic index values. cE(w)s with low Sérsic index values are found to have intermediate ages ($\sim 9.5$ Gyr in median), metal-rich ([Z/H] $\sim 0.14$), and low [$\alpha$/Fe] ($\sim 0.12$) stellar populations, indicating extended SF histories similar to those of pseudo-bulges. In contrast, cE(w)s with high Sérsic index values are characterized by old ($\sim 13$ Gyr), metal-poor ([Z/H] $\sim -0.13$), and high [$\alpha$/Fe] ($\sim 0.32$) populations, reflecting early and rapid SF histories akin to classical bulge properties.

It has been suggested that cE(w)s are remnants of massive progenitors with bulges whose outer envelopes have been tidally stripped by a nearby massive host galaxy (i.e., stripped cEs; see \citealt{Ferre-Mateu18, Kim20, Chen22}). Various progenitor types for stripped cEs have been proposed based on information from the stellar populations, SF histories, and internal kinematics of cEs \citep[see][for details]{Ferre-Mateu18, Ferre-Mateu21a}. Our findings suggest that the progenitors of cE(w)s with low Sérsic index values are likely spiral galaxies with pseudo-bulges exhibiting bursty and extended SF histories. Conversely, cE(w)s with high Sérsic index values may originate from spiral galaxies with classical bulges displaying early and peaked SF histories. Notably, some cE(w)s exhibit high Sérsic index values ($n > 3$) similar to those of cE(w/o)s, which are comparable to those of early-type galaxies \citep[e.g.,][]{Kormendy96, Kormendy09}. This suggests that the progenitors of these cE(w)s might be massive early-type galaxies. Consequently, by incorporating additional information, including stellar populations, we propose that the Sérsic index can serve as a tool to identify the plausible progenitor type of stripped cEs.

\subsection{Double-component Formation of Isolated cEs}

One of the most notable findings of this study is that cE(w/o)s, which do not have a nearby host galaxy, display a structural configuration distinct from that of host-associated cE(w)s. Tidal stripping by a massive host galaxy can eliminate the outer stellar component of the massive progenitors of cE(w)s, resulting in a compact spheroidal core and thus a single-component structure. In contrast, most cE(w/o)s exhibit an additional extended faint stellar component surrounding the central core. The inner component dominates the core’s luminosity, whereas the outer component presents a shallow light distribution, consistent with a diffuse disk or a stellar envelope. \citet{Paudel14} also reported that an isolated cE, CGCG 036-042, exhibits a multi-component structure with a faint stellar envelope extending to 10~kpc from the galaxy center. Our findings suggest that distinct formation processes may be at play in isolated cEs, facilitating the development of more complex structures.

According to in-situ formation scenarios, isolated cEs are inherently low-mass ($\sim 10^8$ -- $10^{10}\,M_\odot$) compact galaxies that formed in the early universe \citep{Wirth84, Kormendy09, Kormendy12}, where they emerged from the mergers of smaller galaxies \citep{Kormendy09, Paudel14, Deeley23}. Although low-mass ($< 10^{10}\ M_{\odot}$) galaxies undergo fewer mergers in galaxy clusters \citep{Lucia06}, both observations and numerical simulations have suggested that substantial merger processes can occur in low-density environments \citep{Klimentowski10, Hopkins09a, Paudel18}. Furthermore, mergers between small galaxies at high redshifts are more prevalent than in the present day \citep[e.g.,][]{Kobayashi04, Hopkins09a, Klimentowski10, Wellons16}. If these mergers are gas-rich, a dissipative phase can form a dense central component, while residual gas settles into a disk-like structure \citep{Mihos96, Robertson06, Hopkins09b, Wuyts10}. The resultant double-component configurations are similar to those identified in our cE(w/o)s. Furthermore, the relatively low Sérsic indices (median value of 1) for the outer components of cE(w/o)s are indicative of disk profiles, thereby supporting the merger-dominated formation pathway. Simulations have also demonstrated that such compact remnants may remain relatively unaffected by other galaxies over cosmic time, persisting as isolated cEs in the local universe \citep{Zolotov15, Wellons16, Deeley23}.

Notably, cEs may exhibit properties akin to massive compact galaxies at high redshifts, including similarities in mass–size relation, stellar population, and kinematics \citep{Yildrim17, Ferre-Mateu21a, Grebol-Tomas23}. Observations of galaxies at high redshifts have revealed that massive, quiescent compact galaxies, referred to as ``red nuggets'', already exist at $z > 2$ (\citealt{Dokkum08}; \citealt{Damjanov14}; \citealt{Schreiber18}; \citealt{Martin-Navarro19}; \citealt{Valentino20}; see also \citealt{Carnall23a, Carnall23b} for red nuggets at $z > 5$ from the JWST observations). A growing body of evidence indicates that, although these galaxies are very compact \citep[]{Zirm07, Buitrago08, Dokkum08, Szomoru10}, a significant proportion exhibit more flattened and disk-like morphologies \citep[]{Toft05, Buitrago08, Buitrago18, van11, Chang13, Zolotov15, Trujillo06, Trujillo14}. Several studies have revealed rotation-dominated kinematics in these galaxies, where such measurements are available, further confirming their disk-like nature from a dynamical perspective \citep{Belli17, Toft17, Newman18}. The analysis of simulated merger remnants also suggests that the disk-like morphology of high-redshift red nuggets is shaped by gas-rich mergers between galaxies \citep{Wuyts10, Wellons15}. Moreover, a non-negligible fraction of early massive compact galaxies is anticipated to survive in the present universe \citep[e.g.,][]{Wellons16, Martinovic17}. These untouched galaxies, dubbed massive relic galaxies in the local universe, also retain disk features \citep[]{Trujillo14, Ferre-Mateu17, Spiniello21}. The double-component outcome observed in the red nuggets and massive relic galaxies bears a remarkable resemblance to that found in local cE(w/o)s, albeit on a smaller mass scale. By analogy, isolated cEs may serve as low-mass counterparts to these red nuggets, having similarly formed in the early universe through gas-rich mergers and subsequently avoiding further significant growth.

\subsection{Environmental Capture and Ejection in cE Formation}

\begin{figure}[h]
\centering
\centering
\includegraphics[width=\textwidth]{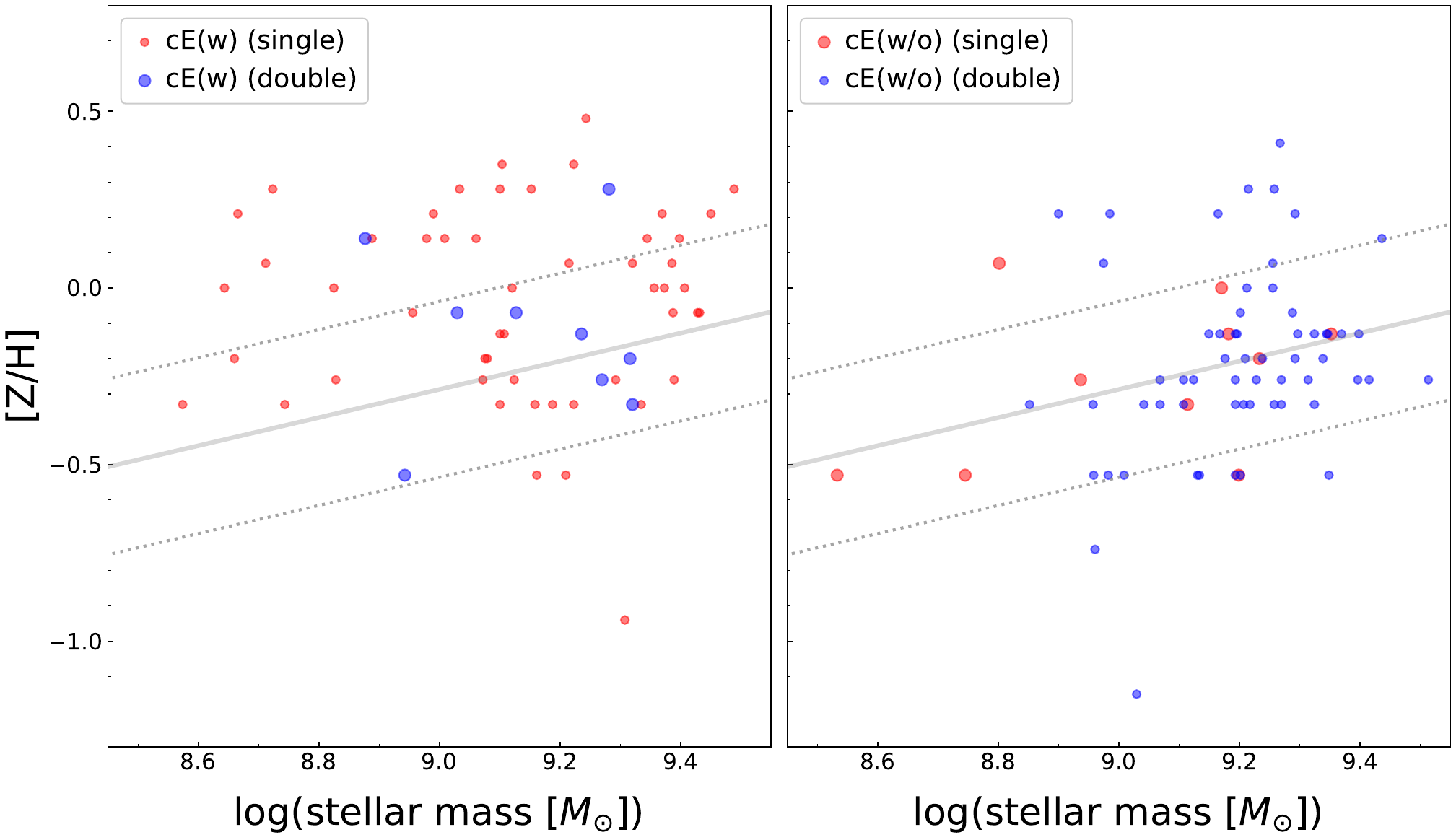}
\caption{Mass–metallicity distribution of cE(w)s (left panel) and cE(w/o)s (right panel). In each panel, red circles represent single-component cEs, while blue circles indicate double-component cEs. The solid line illustrates the linear best fit for cE(w/o)s, with dashed lines indicating the $\pm 1\sigma$ deviation.}
\label{fig:5}
\end{figure}

The mass--metallicity relation of the cEs serves as a diagnostic diagram to discriminate between their different origins. If cEs are remnants of larger progenitors with bulges formed through a stripping process by a neighboring massive host galaxy, they would exhibit higher metallicities compared to galaxies of similar low masses, thus appearing as outliers in the local mass--metallicity relation \citep{Chilingarian15, Janz16, Ferre-Mateu18, Ferre-Mateu21a, Kim20}. In contrast, cEs that follow the mass--metallicity relation are intrinsically low-mass cEs. According to \citet{Kim20}, cE(w)s deviate from the mass--metallicity relation of early-type galaxies, whereas cE(w/o)s conform to this relation. This finding supports the hypothesis that cEs comprise galaxies with two distinct origins based on their local environment: stripped cEs for cE(w) and intrinsic cEs for cE(w/o)s.

In Figure~\ref{fig:5}, we present the mass--metallicity distribution of 61 cE(w)s (left panel) and 70 cE(w/o)s (right panel), along with the linear best fit for cE(w/o)s (solid line) and its $\pm1\sigma$ deviation (dotted lines) as a reference. A majority (74\%, 45/61) of cE(w)s lie above this linear best-fit line, indicating that they have systematically higher [Z/H] values than the mean distribution of cE(w/o)s. If cE(w)s are indeed stripped cEs, they should deviate from the mass--metallicity relation of cE(w/o)s by being more metal-rich at a given mass. In alignment with this scenario, the majority of cE(w)s are more accurately characterized by a single-component structure, which supports tidal stripping as their primary formation mechanism. Nevertheless, approximately 15\% (9/61) of cE(w)s exhibit a double-component structure, suggesting that an alternative channel may contribute to their formation.

A recent investigation utilizing IllustrisTNG-50 cosmological simulations reveals that about 68\% of cEs originate in isolated environments \citep{Deeley23}. However, some of these cEs eventually enter galaxy groups or clusters; while most remain on wider orbits, a few are drawn in and captured by more massive galaxies. Consequently, although the majority of cE(w)s are remnants of massive galaxies that have undergone stripping, some are inherently formed in isolation and subsequently fall into the gravitational potential of a massive host galaxy (see also \citealt{Jang24}). In this context, these cE(w)s might exhibit a double-component structure, rendering them structurally distinct from stripped cEs.

In comparison with the single-component cE(w)s (red circles), the double-component cE(w)s (blue circles) in the left panel appear to exhibit relatively lower [Z/H] values at a given stellar mass, positioning them, on average, below the main distribution of their single-component counterparts. Notably, these double-component cE(w)s also reside near the solid line representing the mean distribution of cE(w/o)s in the right panel, indicating that they align more closely with cE(w/o)s. This feature suggests that some double-component cE(w)s may not have been formed exclusively through the canonical tidal stripping pathway. Instead, they may have formed in isolation and were recently captured by more massive galaxies. This process would allow them to retain the structural and chemical characteristics inherited from their isolation phase, despite now existing in local environment similar to those of the single-component cE(w) population.

Alternatively, a subset of these systems may result from projection effects in the classification of cE(w)s. Given that our classification relies on projected separation and line-of-sight velocity offset, it is plausible that some cEs identified as cE(w)s are not actually associated with the nearby bright galaxy. Instead, they may be genuine cE(w/o)s that have been misclassified due to small projected distances. Although such misclassification is expected to be minor, it cannot be entirely dismissed and may partially account for some cE(w)s exhibiting double-component structures.

Over extended timescales, the outer component of double-component cE(w)s is anticipated to diminish or be entirely stripped by the tidal forces exerted by a host galaxy, resulting in the appearance of single-component cEs. Moreover, if interactions with the host galaxy do not induce additional SF, thereby failing to enhance the overall metallicity, these older captured cEs can effectively maintain their lower metallicities. Consequently, single-component cE(w)s situated below the mean distribution of cE(w/o)s (i.e., red circles below the solid line) may represent systems that were captured at earlier epochs but have subsequently lost any morphological evidence of their former two-component structure.

In the right panel of Figure~\ref{fig:5}, while the majority of cE(w/o)s exhibit a double-component structure (blue circles), 14\% (10/70) of cE(w/o)s appear as single-component systems (red circles). This finding suggests the presence of additional formation pathways for cE(w/o)s, beyond the primary mechanism of intrinsic formation of low-mass cEs. \citet{Chilingarian15} proposed that certain cEs residing in nominally isolated environments may have originated in clusters or groups and were subsequently ejected through three-body encounters. If ejection occurs after the main tidal stripping phase of the cE, during which the progenitor undergoes tidal stripping by the massive host galaxy, it would already exhibit a compact, single-component structure akin to cE(w)s.

An alternative formation channel of single-component cE(w/o)s emerges from the possibility that certain cEs in the cluster or group environments are intrinsically formed. \citet{Martinovic17} demonstrated that cEs can form in-situ within dense environments and subsequently become satellite galaxies of more massive galaxies. Over time, tidal interactions with a host galaxy may eliminate any extended outer component, leading to a single-component structure. Under such conditions, SF can remain suppressed, limiting further metallicity enrichment and preserving relatively low [Z/H] values. Some of these cEs may also be ejected from clusters or groups through three-body encounters \citep{Chilingarian15}. Consequently, these cEs are candidates for single-component cE(w/o)s found around the mass--metallicity relation of cE(w/o)s. In both scenarios, the ejection of tidally stripped cEs and intrinsic cEs in dense environments underscores the diversity of formation and evolution pathways that can produce single-component cE(w/o)s in isolated environments.

\section{SUMMARY and CONCLUSIONS}
\label{sec:summary, conclusion}

In this work, we investigated the internal structures of 138 local cEs at $z < 0.05$ by performing single- and double-component Sérsic model fitting to their SDSS $r$-band SBPs. By classifying cEs into those located near a massive host galaxy (cE(w)s) and those in isolation without a nearby host (cE(w/o)s), we investigated whether distinct local environment influence structural differences. Our main findings are as follows:

\begin{enumerate}
    \item Most host-associated cE(w)s ($\sim$85\%) display a single-component Sérsic profiles, with no indication of extended outer envelopes. In contrast, the majority ($\sim$85\%) of cE(w/o)s exhibit double-component Sérsic profiles, consisting of a compact, high-surface-brightness inner component and a diffuse, disk-like outer component. These results strongly suggest that the local galactic environment, particularly the presence or absence of a host galaxy, plays a crucial role in shaping the structures of cEs.
    
    \item The single-component nature and high metallicities of cE(w)s can be attributed to the stripping of a large, massive progenitor’s outer envelope, resulting in a compact bulge-like remnant. The presence of a large, faint stellar envelope in most cE(w/o)s indicates gas-rich mergers at high redshifts, leading to the formation of dense central cores surrounded by more diffuse disks. These intrinsically formed cE(w/o)s may serve as low-mass analogs to the massive ``red nuggets,'' which formed in the early universe and subsequently avoided significant growth.
    
    \item Many cE(w)s have lower Sérsic index values ($n < 2$), indicative of pseudo-bulge progenitors, while others have higher indices ($n > 2$), suggesting origins from classical-bulge or elliptical-like progenitors. The properties of stellar populations further corroborate this dichotomy: cE(w)s with lower Sérsic indices have relatively younger, more metal-rich, and low $\alpha$ element stellar populations, resembling pseudo-bulges, whereas those with higher indices are characterized by older, metal-poor, and  $\alpha$-enhanced stellar populations akin to classical bulges. Thus, the Sérsic index is particularly effective in distinguishing the progenitor nature of stripped cEs. 
    
    \item A minor subset of cE(w)s exhibits double-component structures and lower metallicities similar to cE(w/o)s. This implies that some isolated cEs, which form intrinsically, might later be ``captured'' by a massive host galaxy, while preserving their structural and chemical characteristics inherited from their isolated phase. Alternatively, some of these objects may be the result of projection effects in the classification of the local environment, leading to the misidentification of genuine cE(w/o)s. A few cE(w/o)s with single-component profiles could result from tidal stripping or intrinsic formation within clusters and groups, eventually becoming ``ejected'' systems from these dense environments. These objects underscore the diversity of formation and evolutionary pathways that can produce single- or double-component cEs in either environment.
\end{enumerate}

In summary, our study emphasizes that the internal structures of cEs, particularly the presence or absence of extended outer envelopes, provide critical insights into disentangling formation pathways, ranging from remnants of tidally stripped massive galaxies to survivors of early, gas-rich mergers. High-quality imaging from ground-based or space-based telescopes will help refine the structures of cEs by minimizing uncertainties associated with low-surface-brightness structures. Moreover, both existing and additional spectroscopy can provide more precise constraints on the internal kinematics and SF histories of cEs, thereby enhancing our ability to differentiate their diverse formation and evolutionary processes.

\section*{}

We are grateful to the anonymous referee for helpful comments and suggestions that improved the clarity and quality of this paper. S.C.R. contributed to this work as a corresponding author. This work was supported by the National Research Foundation of Korea through grants, NRF-2022R1A2C1007721 (S.C.R.), NRF-2019R1I1A1A01061237 and NRF-2022R1C1C2005539 (S.K.), and NRF-2022R1I1A1A01054555 (Y.L.).

\bibliographystyle{aasjournal} 
\bibliography{reference.bib}

\end{document}